\documentclass[10pt,twocolumn,letterpaper]{IEEEtran}

\def\ps@headings{%
\def\@oddhead{\mbox{}\scriptsize\rightmark \hfil \thepage}%
\def\@evenhead{\scriptsize\thepage \hfil \leftmark\mbox{}}%
\def\@oddfoot{}%
\def\@evenfoot{}}

\IEEEoverridecommandlockouts

\usepackage[utf8]{inputenc}
\usepackage{amsmath}
\usepackage{amsfonts}
\usepackage{amssymb}	
\usepackage{graphicx}
\usepackage{xcolor}
\usepackage{hyperref}
\usepackage{comment}
\usepackage{multirow}
\usepackage{subcaption}
\usepackage{soul}

\DeclareMathOperator*{\E}{\mathbb{E}}

\usepackage{subcaption}

\definecolor{boristext}{rgb}{0.22, 0.44, 0.88}
\definecolor{boriscomments}{rgb}{0.88, 0.04, 0.04}
\definecolor{boristochange}{rgb}{0.2, 0.8, 0.8}

\begin{document}

\title{On the Low-latency Region of Best-effort Links \\for Delay-Sensitive Streaming Traffic}

\author{Boris Bellalta\thanks{B. Bellalta (\textit{boris.bellalta@upf.edu}) is  with  Universitat Pompeu Fabra, Barcelona. The author would like to thank Marc Carrascosa for his contribution in taking the Google Stadia measurements. Also, the author wants to especially thank the anonymous reviewers for their constructive, insightful and challenging comments. This work was supported by WINDMAL PGC2018-099959-B-I00 (MCIU/AEI/FEDER,UE), and SGR017-1188 (AGAUR). }}
\date{}

\maketitle

\begin{abstract}
This letter analyzes the Low-latency Region (LLR) of a best-effort link (i.e., no traffic differentiation, and first come first serve scheduling) carrying both delay-sensitive (DS) streaming and non-delay-sensitive (NDS) background traffic. Moreover, inside the LLR, we show it exists a proportional fair arrival rate allocation for both the DS and NDS traffic streams. This optimal operating point results from maximizing a simple throughput-delay trade-off that considers the NDS traffic load, and the mean delay of the DS packets. To show how the presented trade-off could be used to allocate NDS traffic in a realistic scenario, we use Google Stadia traffic traces to generate the DS flow. Results from this use-case confirm that the throughput-delay trade-off also works regardless the distribution of the packet arrival and packet service times. 
\end{abstract}



\section{Introduction}

The success of delay-sensitive (DS) audio and video streaming services such as voice over IP, real-time videoconferencing, cloud-gaming and virtual reality in best-effort networks depends to a great extent on the network's ability to guarantee low end-to-end delays. This is especially challenging since, with the aim to keep the network neutrality, no traffic differentiation is usually provided between DS and non-delay-sensitive (NDS) background traffic, and packets are transmitted following a first come first serve scheduling in the multiple links they traverse from the source to the destination.

The interactions between DS streaming and NDS background traffic have been widely studied in the literature. In the one hand, the presence of NDS background streams results in higher delays and packet losses for the DS streams \cite{de2011skype,carlucci2018controlling}. However, in the other hand, NDS background traffic may also suffer from starvation in presence of DS streams. To mitigate this situation, DS streams also implement rate adaptation and congestion control solutions, such as the Google Congestion Control algorithm used in WebRTC~\cite{carlucci2017congestion,alos2019congestion}. 

In the described context, we aim to find how much DS and NDS traffic can be allocated to a best-effort link without compromising the delay requirements of the DS traffic. The set of those traffic loads define what we call the Low-latency region (LLR). With respect to the delay requirements of a DS stream, we use the following criterion: \textit{a DS stream operates inside the low-latency region if the average time that a DS packet spends in the link is lower than the mean inter-DS packet arrival time}. Note that this is equivalent to say, that a new arriving packet must leave the system, on average, before another packet of the same flow arrives to it.

Using the aforementioned delay criterion, and modelling a best-effort link as a M/G/1 queue, we first calculate how much DS streaming and NDS background traffic can be allocated to the link. This gives us the upper limit of the LLR. Then,  we show that inside the LLR it exists a proportional fair rate allocation for both DS and NDS traffic. This fair rate allocation corresponds to the point at which the difference between the throughput gain when NDS background traffic is added to the link, and the corresponding latency loss for the DS traffic is maximum. We will refer to this point as the proportional fair low-latency (PFLL) rate allocation.

The PFLL rate allocation can be achieved in practice by estimating the mean time between DS packet arrivals, the mean packet delay of the DS packets, and controlling the amount of NDS traffic that can be allocated to a link. It requires, however, to be able to identify and classify the active traffic streams, which can be done using Machine Learning techniques~\cite{nguyen2008survey}. Once traffic streams are identified and classified, traffic shaping techniques can be used to control the amount of traffic that is allowed to enter the link, relying in the traffic source for adapting the traffic generation rate. 

The use of the mean packet delay as a criterion to define the LLR instead of a certain percentile results in simple, closed-form, and insightful expressions. However, and more importantly, the use of the mean offers by itself an interesting trade-off between the carried load and the packet delay of a link in absence of any information regarding the delay requirements of the DS streaming traffic, neither on how many other links are traversed by the DS flow between its source and destination. The use of the mean, and this paper in general, was partially inspired by \cite{gringoli2019regulating}, where the authors play with the packet aggregation level in a Wi-Fi network to guarantee both high-throughput and low-latency.

Finally, we study the PFLL rate allocation when a link carries Google Stadia traffic \cite{carrascosa2020cloud}, a cloud-gaming service that requires both high-throughput and low latency to perform satisfactorily. We aim to obtain how much NDS traffic can be allocated, and evaluate how much it disturbs the DS traffic stream in terms of the extra added latency. However, in addition to that, and more importantly, we confirm that we can estimate the PFLL rate allocation even if the traffic arrival process is not Poisson.


\section{System Model} \label{Sec:SystemModel}

\begin{figure}[t!]
\centering
\includegraphics[width=0.75\columnwidth]{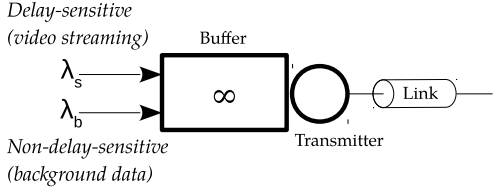}
\caption{Link model. The buffer and the transmitter correspond to the network interface.} 
\label{Fig:FP_problem}
\end{figure}

We consider that a link consists of a buffer and a transmitter as shown in Fig.~\ref{Fig:FP_problem}. The link capacity, i.e., the rate at which the transmitter works, is of $R$~bits/second. We assume the buffer is large enough to be considered of infinite size. We also assume all arriving flows have the same priority. Therefore, traffic differentiation is not applied, and all arriving packets are served following their order of arrival.

Besides, to keep the analysis simple, we do the following considerations:\footnote{In Sec. \ref{Sec:Stadia}, we study a case where packets do not arrive following a Poisson process, and the DS and NDS flows have different service distributions.} 
\begin{enumerate}
	\item There are only one DS flow and one NDS flow. The packet arrival process for both flows follows a Poisson process, with mean rates $\lambda_s$ and $\lambda_b$ packets/second, respectively. The mean aggregate packet arrival rate is given by $\lambda=\lambda_s+\lambda_{b}$. 
	\item The two flows have the same service distribution, with $S$ the random variable representing the packet service time. Service times are independent and identically distributed. The service time distribution is characterized by its mean, $\E[S]$, and its coefficient of variation $C_S$. The mean transmission rate of the link in packets/sec. is $\mu=1/\E[S]$. 
\end{enumerate}

Taking into account these considerations, the link is modelled as an M/G/1 queueing system \cite{bertsekas1992data}. The mean sojourn time for a packet of flow~$s$, $\E[D_s(\lambda_s,\lambda_b)]$ and for a packet of flow~$b$, $\E[D_b(\lambda_s,\lambda_b)]$, is the same: 
\begin{align} \label{Eq:MG1_ED}
	\E[D_s(\lambda_s,\lambda_b)]&=\E[D_b(\lambda_s,\lambda_b)]=\E[S]+\frac{\lambda \E[S^2]}{2(1-a)} \nonumber \\
	&=\frac{1}{\mu}\left(1+\frac{a}{1-a}\theta\right) =\frac{\Gamma(\lambda,\theta)}{\mu-\lambda},
\end{align}
where $\E[S^2]=\E^2[S](1+C_S^2)$ is the second moment of the service time, $a=\lambda/\mu$ is the link utilization, and $\theta={(1+C^2_{S})}/{2}$. Lastly, $\Gamma(\lambda,\theta)=1-a+a\theta$ is the ratio between the delay of a M/G/1 and a M/M/1 queue. 
Note that $a=a_s+a_b=\lambda_s/\mu+\lambda_b/\mu$, where $a_s$ and $a_b$ are the link utilization by the DS and NDS flows, respectively.

Note that the presented system model can represent either a point-to-point link of a network (e.g., the network interface of a router/switch) or a point-to-multipoint link (e.g., a WiFi Access Point with several associated clients). The only requirement is that the DS and NDS flows share the same buffer and transmitter.


\section{Low-latency Region}

Following our previous definition, in the absence of NDS traffic~($\lambda_b=0$), a DS flow $s$ is working in the low-latency region of a link if the following condition is satisfied:
\begin{align}\label{Eq:ConditionLLRegion}
	\frac{1}{\lambda_s} \geq \E[D_{s}(\lambda_s,0)] =\frac{\Gamma(\lambda_s,\theta)}{\mu-\lambda_s}.
\end{align}

The highest value of $\lambda_s$ that satisfies \eqref{Eq:ConditionLLRegion} is
\begin{align} 	\label{Eq:LowLatencyRegion}
	\lambda_s^{+} = \frac{\mu}{1+\sqrt{\theta}},
\end{align}
and therefore, the LLR includes all  $\lambda_s \in [0,\lambda_s^+]$. Note that the value of $\lambda_s^{+}$ depends on the service time distribution of the DS traffic. For example, if the service time distribution is exponential, $\lambda^+_s$ is half of the link capacity.
 
\begin{figure}[t!]
\centering
\includegraphics[width=0.9\columnwidth]{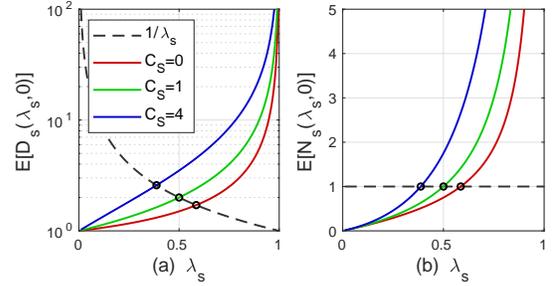}
\caption{(a) Mean DS packet delay and, (b) Mean number of DS packets in the link for different $C_S$ values. The $\lambda^+_s$ values are indicated using black circles. The plots are obtained for $\mu=1$ packets/second.}
\label{Fig:FP_LL_region}
\end{figure}

Considering Little's theorem, (\ref{Eq:ConditionLLRegion}) can be rewritten as:
\begin{align}
	\lambda_s\E[D_s(\lambda_s,0)] = \E[N_s(\lambda_s,0)]\leq 1, \nonumber
\end{align}
where $\E[N_s(\lambda_s,0)]$ is the mean number of DS packets in the link. We can observe that on average, the DS flow will be inside the LLR if the mean number of packets in the link is equal or less than one.

Fig. \ref{Fig:FP_LL_region}.a) shows the values of $\lambda_s^{+}$ as the intersection between $1/\lambda_s$ and $\E[D_s(\lambda_s,0)]$ for different service time distributions. As expected, higher $C_S$ values reduce the low-latency region. Fig.~\ref{Fig:FP_LL_region}.b) confirms that $\E[N_s(\lambda_s,0)] \leq 1$ when working inside the LLR. 


\section{Low-latency Rate Allocation}

Let us consider now that an NDS flow $b$ is added to the link, and so it shares the link resources with the DS flow~$s$. We are then interested in finding how much NDS traffic can be admitted to the link while keeping the DS flow working inside the LLR.

In the following, we assume $\lambda_s \leq \lambda_s^+$ is known, and therefore, we focus on obtaining the value of $\lambda_b$. The other way around, i.e., given a certain amount of NDS traffic in a link, to obtain how much DS traffic can be allocated, gives exactly the same results, as expected.

\subsection{Max low-latency rate allocation}

A solution to the previous problem can be found by solving 
\begin{align}
	\frac{1}{\lambda_s} \geq \E[D_s(\lambda_s,\lambda_b)]= \frac{\Gamma(\lambda,\theta)}{\mu-\lambda}. \nonumber
\end{align}

Then, we can obtain $\lambda_b^+$ as the highest feasible NDS rate allocation, i.e.,
\begin{align} \label{Eq:MaxThroughput}
	 \lambda_b^+ = \left(\frac{\mu-\lambda_s}{\Gamma(\lambda_s,\theta)}\right)-\lambda_s=\frac{1}{\E[D_s(\lambda_s,0)]}-\lambda_s, 
\end{align}
which explicitly depends on the mean delay of the DS packets. 

We can re-write \eqref{Eq:MaxThroughput} as $\lambda_b^+ = \mu - \kappa^{+} \lambda_s$, with $\kappa^+\lambda_s$ representing the minimum link capacity required by the DS flow to work in the LLR. The value of $\kappa^+$ is obtained by developing the term $\Gamma(\lambda_s,\theta)$ as follows:
\begin{align}
	\lambda_b^+ &= \left(\frac{\mu-\lambda_s}{1-\frac{1}{\mu}(\lambda_s-\lambda_s\theta)}\right)-\lambda_s = \mu  \underbrace{\left(\frac{\mu-\lambda_s}{\mu-(\lambda_s-\lambda_s\theta)}\right)}_{=\alpha}  - \lambda_s. \nonumber 
\end{align}
Then, we obtain $\kappa^+$ by considering that $\lambda_b^+ = \mu\alpha - \lambda_s = \mu - \kappa^+\lambda_s$. Therefore,
\begin{align}
	\kappa^+ &= \frac{\mu}{\lambda_s}(1-\alpha)+1 = \frac{\mu}{\lambda_s}\left(\frac{\lambda_s\theta}{\mu-\lambda_s+\lambda_s\theta}\right)+1  \nonumber \\ 
			 &=	 1 +  \left(\frac{\theta}{\frac{\mu-\lambda_s+\lambda_s\theta}{\mu}}\right) = 1 + \frac{\theta}{\Gamma(\lambda_s,\theta)}. \nonumber		 
\end{align}
Let us now define $\beta$ to refer to the second term of $\kappa^+$, i.e., $\beta=\frac{\theta}{\Gamma(\lambda_s,\theta)}$. Note that $\beta\lambda_s$ is the amount of link capacity that has to remain unused to keep the DS flow operating inside the LLR. For example, considering the service time is exponentially distributed, we have that $\kappa^+=2$, and $\lambda^+_{b} = \mu - 2\lambda_s$. In this case, the amount of unused link capacity is equal to the load of the DS flow, as~$\beta=1$.

Finally, observe that $\kappa^+ \lambda_s^+ = \mu$. This result could be expected since for $\lambda_s^+$ we have that $\lambda_b=0$. In detail, by expanding the terms $\kappa^+$ and $\lambda_s^+$, we have that
\begin{align}
	\left(1 + \frac{\theta}{\Gamma(\lambda^+_s,\theta)} \right) \frac{\mu}{1+\sqrt{\theta}}=\mu, \nonumber  
\end{align}
where 
\begin{align}
	\frac{\theta}{\Gamma(\lambda^+_s,\theta)} = \frac{\theta}{1-\frac{1}{1+\sqrt{\theta}}+\frac{\theta}{1+\sqrt{\theta}} }=\frac{\theta(1+\sqrt{\theta})}{\theta+\sqrt{\theta}}=\sqrt{\theta}. \nonumber
\end{align}

This last result tells us that when we are operating in the limit of the low-latency region, the amount of idle bandwidth is equal to $\lambda^+_s \sqrt{\theta}$, which further shows the relationship between the service time distribution and the LLR.


\subsection{Proportional fair low-latency rate allocation}

Let $\lambda^*_b \leq \lambda^+_b$ be the proportional fair low-latency rate allocation, i.e., the value at which the trade-off between the delay of DS packets and the NDS throughput is maximum. $\lambda_b > \lambda^*_b$ values result in a higher delay increase for DS packets than the link throughput gain. Similarly, for $\lambda_b < \lambda^*_b$ values, we observe the opposite result.

To find $\lambda^*_b$, we formulate the throughput-delay trade-off as the difference between the link throughput gain, and the delay loss for the DS traffic, 
\begin{align} \label{Eq:gainfunction}
	g(\lambda_s,\lambda_{b})=G_T(\lambda_s,\lambda_{b})-G_D(\lambda_s,\lambda_{b}),
\end{align}
with respect to the case there is no NDS traffic.
 
The NDS throughput gain is 
\begin{align}
	G_{T}(\lambda_s,\lambda_{b}) = \frac{(\lambda_b+\lambda_s)-\lambda_{s}}{\lambda_{s}}=\frac{\lambda_b}{\lambda_s}, \nonumber
\end{align}
and the DS delay loss is 
\begin{align}
G_{D}(\lambda_s,\lambda_{b})&= \frac{\E[D_s(\lambda_s,\lambda_b)]-\E[D_s(\lambda_s,0)]}{\E[D_s(\lambda_s,0)]} \nonumber \\
&=\frac{\frac{1}{\mu}+\frac{1}{\mu}\frac{a}{(1-a)}\theta-\frac{1}{\mu}-\frac{1}{\mu}\frac{a_s}{(1-a_s)}\theta}{\frac{1}{\mu}+\frac{1}{\mu}\frac{a_s}{(1-a_s)}\theta}\nonumber \\
& = \frac{\beta(\lambda-\lambda_s)}{\mu-\lambda} =  \frac{\beta \lambda_b}{\mu-\lambda_s-\lambda_b}. \nonumber
\end{align}

Then, \eqref{Eq:gainfunction} results in 
\begin{align}\label{Eq:g}
	g(\lambda_s,\lambda_b)&= \frac{\lambda_b}{\lambda_s} - \frac{\beta \lambda_b}{\mu-\lambda_s-\lambda_b}  = \lambda_b\left(\frac{1}{\lambda_s}-\frac{\beta}{\mu-\lambda}\right). 
\end{align}

Finally, we are interested in finding
\begin{align}
	\lambda^{*}_b = \underset{\lambda_b}{\operatorname{argmax}}~g(\lambda_s,\lambda_{b}), \nonumber 
\end{align}
that is the proportional fair low-latency rate allocation.\footnote{Note that just taking logs in \eqref{Eq:g} we get $\log(\lambda_b)+\log\left(\frac{1}{\lambda_s}-\frac{\beta}{\mu-\lambda} \right)$, which satisfies the definition of proportional fairness, i.e., the gain (increase) in one variable should be higher than the loss (decrease) in the others, which is achieved by maximizing the sum of the logarithms of the variables under study.}

Since $g(\lambda_s,\lambda_b)$ is concave, and has its maximum in the range $\lambda_b \in [0,\lambda^+_b$], we find $\lambda^{*}_b$ by deriving \eqref{Eq:g} with respect to~$\lambda_b$. The first derivative of $g(\lambda_s,\lambda_b)$ is
\begin{align}
	\frac{d g(\lambda_s,\lambda_b)}{d\lambda_b}=\frac{1}{\lambda_s} - \frac{\beta (\mu-\lambda_s)}{(\mu-\lambda_s-\lambda_b)^2},  \nonumber
\end{align}
and it is equal to zero for
\begin{align}
\lambda^*_b & = \mu - \lambda_s -  \sqrt{\beta\lambda_s(\mu-\lambda_s)}  \nonumber \\ 
& = \mu - \lambda_s\left(1 +  \sqrt{\beta \left(\frac{\mu-\lambda_s}{\lambda_s}\right)} \right) = \mu -\kappa^* \lambda_s,
\end{align}
with $\kappa^* = 1+ \sqrt{\beta \left(\frac{\mu-\lambda_s}{\lambda_s}\right)}$. Similarly to the max NDS rate allocation, the $\kappa^* \lambda_s$ term is the link capacity required by the DS flow to work at the PFLL rate allocation. Note that $\kappa^* \geq \kappa^+$ for $\lambda_s \leq \lambda^+_s$.
 
\begin{figure}[t!]
	\centering
	\includegraphics[width=0.90\columnwidth]{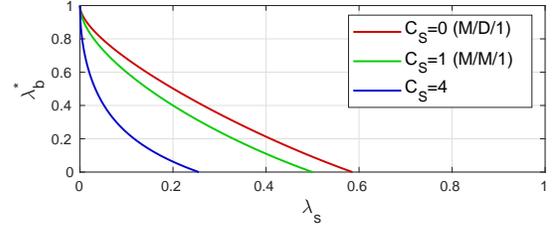}
	\caption{Proportional Fair Low-latency rate allocation for different $C_S$ values and $\mu=1$.}
	\label{Fig:LLRegion}
\end{figure}

Fig. \ref{Fig:LLRegion} shows the PFLL rate allocation ($\lambda^*_b$ values) for different $\lambda_s$ and $C_S$ values. For low values of $\lambda_s$, almost full link capacity can be achieved. However, it rapidly reduces when $\lambda_s$ increases. Also, increasing the variability of the service time distribution reduces the PFLL rate allocation.


\subsection{Max vs PFLL rate allocation}

\begin{figure}[t!]
	\centering
	\includegraphics[width=0.90\columnwidth]{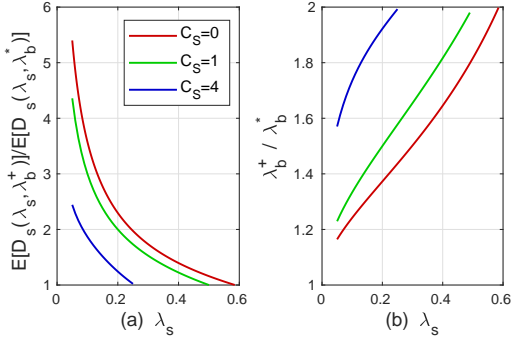}
	\caption{(a) Delay reduction vs (b) Throughput increase between max and PFLL rate allocation strategies, with $\mu=1$.}
	\label{Fig:ComparativeMaxPF}
\end{figure}

To provide some further insights on the significance of the max and PFLL rate allocation strategies, Fig. \ref{Fig:ComparativeMaxPF} compares $\E[D_s(\lambda_s,\lambda_b^+)]$ and $\E[D_s(\lambda_s,\lambda_b^*)]$, as well as the values of $\lambda_b^+$ and $\lambda_b^*$. It can be observed that the PFLL rate allocation results in a significant delay reduction (i.e., values higher than~$1$ in Fig. 4.a)) for a minimum background traffic loss (values higher than $1$ in Fig. 4.b)) when the load of the DS traffic stream is low. For example, for $C_S=0$ and $\lambda_s\approx 0.1$, the delay is $5$ times lower, while the throughput is reduced only by a factor of $1.2$. Increasing the load of the DS stream reduces the gain in delay and increases the loss in NDS throughput.

\section{An Alternative Formulation}\label{Sec:Alternative}

In this section, we show that the following alternative formulation to \eqref{Eq:g},
\begin{align}\label{Eq:Alternative}
	f(\lambda_s,\lambda_b)&=\lambda_b\left(\E[D_s(\lambda_s,\lambda^+_b)]-\E[D_s(\lambda_s,\lambda_b)]\right),
\end{align}
can be used to accurately estimate the PFLL rate allocation. Note that \eqref{Eq:Alternative} is simply the product between the NDS throughput (the gain), and the difference in delay between the maximum tolerable and the current delay of the DS packets (the loss). To obtain it we have simply approximated the term $\beta/(\mu-\lambda)$ in \eqref{Eq:g} by $E[D_s(\lambda_s,\lambda_b)]$.

\begin{figure}[t]
	\centering
	\includegraphics[width=0.90\columnwidth]{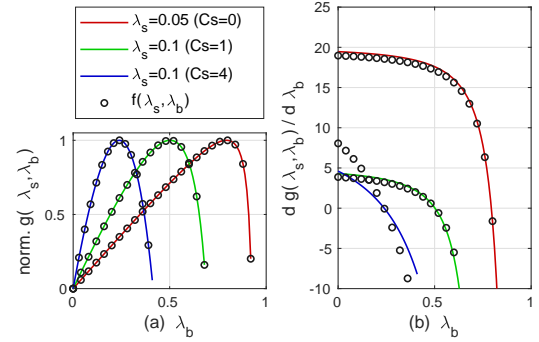}
	\caption{(a) $\hat{g}(\lambda_s,\lambda_b)$ vs $\hat{f}(\lambda_s,\lambda_b)$, and (b) $\frac{dg(\lambda_s,\lambda_b)}{d\lambda_b}$ vs $\frac{df(\lambda_s,\lambda_b)}{d\lambda_b}$ for different $\lambda_s$ and $C_S$ pair of values, with $\mu=1$.}
	\label{Fig:AlternativeGain}
\end{figure}

To evaluate the accuracy of \eqref{Eq:Alternative}, we compare the normalized versions of $\hat{g}(\lambda_s,\lambda_b)=\frac{{g}(\lambda_s,\lambda_b)}{\max({g}(\lambda_s,\lambda_b))}$ and $\hat{f}(\lambda_s,\lambda_b)=\frac{{f}(\lambda_s,\lambda_b)}{\max({f}(\lambda_s,\lambda_b))}$, and the first derivatives of $g(\lambda_s,\lambda_b)$ and $f(\lambda_s,\lambda_b)$. Fig. \ref{Fig:AlternativeGain}.a) shows that both $\hat{g}(\lambda_s,\lambda_b)$ and $\hat{f}(\lambda_s,\lambda_b)$ give the same values. Fig. \ref{Fig:AlternativeGain}.b) shows that even if the first derivatives of $g(\lambda_s,\lambda_b)$ and $f(\lambda_s,\lambda_b)$ are not the same, they are equal to $0$ for the same $\lambda_b$ value. 

\section{Use-case: Google Stadia Traffic} \label{Sec:Stadia}

In this section, we aim to illustrate the existence of the PFLL rate allocation, and the applicability of \eqref{Eq:Alternative}, when the aggregate traffic arrival process is not Poisson, and the DS and NDS traffic flows are characterized by different traffic arrival processes and service time distributions. We also examine the cumulative distribution function (cdf) of the delay of the DS packets, i.e., $D_s(\lambda_s,\lambda_b)$, for different values of $\lambda_b$ to observe how it changes when the NDS background traffic increases.

\begin{figure}[t]
\centering
\includegraphics[width=0.90\columnwidth]{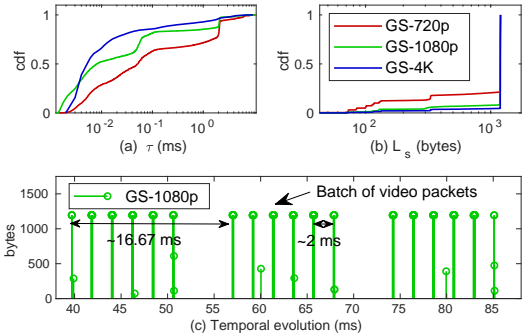}
\caption{Google Stadia's Tomb Raider traffic: cdfs of the (a) inter-packet arrival time ($\tau$), and (b) packet size ($L_s$), and a (c) snapshot of the GS traffic temporal evolution.}
\label{Fig:StadiaCDFs}
\end{figure}

We implemented the system model described in Section \ref{Sec:SystemModel} in C++ using the COST simulation library \cite{chen2002cost}. The DS traffic is generated using a set of traffic traces collected while playing with Google Stadia's Tomb Raider (GS).\footnote{The traces can be found at \href{https://www.upf.edu/web/wnrg/wn-datasets}{https://www.upf.edu/web/wnrg/wn-datasets}.} The traces represent the downlink traffic (mostly video contents, from the server to the client, sent at a rate of 60~frames/second) for the three different video resolutions available in GS (720p, 1080p, and 2160p). The duration of each trace is 30 seconds. Their main characteristics are shown in Fig. \ref{Fig:StadiaCDFs} and in Table~\ref{Tbl:GSStadiaCharacteristics}. We can observe that GS packets arrive in batches of mean size $\E[\sigma]$. In those conditions, packet arrivals are not Poisson, even if the coefficient of variation of the inter-packet arrival time, $C_{\tau}$, for 720p and 1080p is close to~$1$. NDS traffic arrives to the link following a Poisson process. NDS packet sizes are assumed to be exponentially distributed, with an average size of $\E[L_b]=10000$ bits. The capacity of the link is set to $R=100$ Mbps.

\begin{table}[t!]
	\centering
	\begin{tabular}{ccccccc}
		Resolution & Load & $\E[\tau]$ & $C_{\tau}$ & $\E[L_s]$ & $C_{S}$ & $\E[\sigma]$\\		
		 (unit)		   & (Mbps) & (ms) & - & (Bytes) & - & (packets)		\\				
		\hline
		720p (HD) & 10.25 & 1.700 & 0.97 & 997.5 & 0.40 & 2.18  \\
		1080p (FHD) & 27.47 & 1.417 & 0.94 & 1123.2 & 0.23 & 4.33 \\
		2160p (4K)  & 39.89 & 1.293 & 2.87 & 1144.2 & 0.19 & 5.74 
	\end{tabular}
	\caption{Characteristics of Tom Raider Downlink traffic. $\E[L_s]$ is the mean packet size, $\E[\tau]$ is the mean inter-packet arrival time, $\E[\sigma]$ is the mean batch size considering all type of arriving packets, and $C_{\tau}$ and $C_{S}$ are the coefficient of variation of the inter-packet arrival time and service time, respectively.}
	\label{Tbl:GSStadiaCharacteristics}
\end{table}

\begin{figure}[t!]
\centering
\includegraphics[width=0.90\columnwidth]{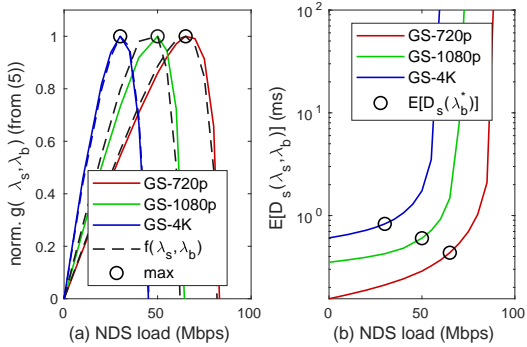}
\caption{(a) $\hat{g}(\lambda_s,\lambda_b)$, and (b) $\E[D_s(\lambda_s,\lambda_b)]$, for different GS video resolutions.}
\label{Fig:Stadia}
\end{figure}

Fig. \ref{Fig:Stadia}.a) shows the value of $\hat{g}(\lambda_s,\lambda_b)$ computed from the simulation data for the three GS video resolutions. For each resolution, we also plot $\hat{f}(\lambda_s,\lambda_b)$, and indicate where the maximum of $\hat{g}(\lambda_s,\lambda_b)$ and  $\hat{f}(\lambda_s,\lambda_b)$ is (circle). We can observe that 1) the PFLL rate allocation exists even if the traffic arrival process is not Poisson, and 2) we confirm~\eqref{Eq:Alternative} is an accurate estimator of the PFLL rate allocation. In Fig. \ref{Fig:Stadia}.b), we plot the mean packet delay for GS packets, $\E[D_s(\lambda_s,\lambda_b)]$, indicating the delay that corresponds to the PFLL rate allocation (circle). The proportional fair NDS rate allocation for the best-effort flow is 65, 50 and 30 Mbps, for 720p, 1080p, and 2160p video resolutions, respectively. Note that in case the DS streaming traffic load changes because GS decides to switch to a different resolution, the PFLL rate allocation does it also accordingly.

\begin{figure}[t!]
\centering
\includegraphics[width=0.90\columnwidth]{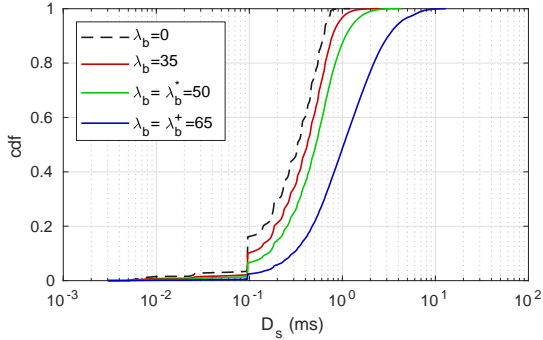}
\caption{cdf of $D_s(\lambda_s,\lambda_b)$ for a GS video resolution of 1080p, and different $\lambda_b$ values.}
\label{Fig:StadiaPercentiles}
\end{figure}

Fig. \ref{Fig:StadiaPercentiles} shows the cdf of the packet delay for GS traffic, $D_s(\lambda_s,\lambda_b)$, for different $\lambda_b$ values. When $\lambda_b > \lambda_b^*$, the negative effect of the NDS traffic on the DS delay is significant. Instead, for $\lambda_b \leq \lambda_b^*$, the cdfs are relatively similar to the case without NDS traffic (black dashed line), hence showing also the benefits of operating close to $\lambda_b^*$. For example, the 90th-percentile of $D_s(\lambda_s,\lambda_b)$ increases by a factor of 1.7x from $\lambda_b=0$ to $\lambda_b=\lambda^*_b$, and by a factor of 4.9x to $\lambda_b=\lambda^+_b$, while the gain in NDS traffic from $\lambda^*_b$ to $\lambda^+_b$ is only of the 30~\%. 

In summary, the presented use-case illustrates how the proposed LLR and rate allocation strategies can be used to characterize the interactions between DS and NDS traffic in best-effort links, providing a framework to better understand the trade-off between capacity and latency in communication networks.


\section{Conclusions}

This paper defines the low-latency region of a best-effort communication link. It also provides two strategies to allocate the rate of NDS background traffic to the link without negatively affecting the DS streaming traffic: the max and PFLL rate allocation strategies. Moreover, the presented analysis shows how the LLR and rate allocation strategies depend on the incoming traffic characteristics (i.e., inter-packet delay, and service time distribution), and DS and NDS traffic loads, providing insights on aspects such as the amount of link capacity that must remain empty to preserve the low-latency operation.

Future work may consider to re-define the low-latency region and rate allocation strategies in terms of delay percentiles, as well as to consider the presence of multiple DS and NDS traffic flows with different arrival and service time distributions. Also, while we have shown that \eqref{Eq:Alternative} works well when the packet arrival process is not Poisson, at least in the considered use-case, we conjecture it may also work when more complex transmission schemes are considered, such as the cases of packet aggregation and multiuser transmissions in WiFi networks. Indeed, we have special interest in applying the results presented in this letter to deal with latency in future WiFi networks \cite{adame2019time}. Moreover, it is also worth to study how the use of the proposed rate allocation strategies could improve the end-to-end latency in multi-hop networks, as well as to investigate the interplay with end-side congestion control techniques.


\bibliographystyle{unsrt}
\bibliography{References}

\end{document}